\documentclass[prb,aps,twocolumn,floatfix]{revtex4}
\usepackage{graphicx}
  \usepackage{amsmath}
  \usepackage{subfigure}
\usepackage{tikz}
\usetikzlibrary{arrows} 
\usepackage{amssymb}
\usepackage{amsfonts}
\usepackage{bm}
\begin{document}

\newcommand {\bea}{\begin{equation}}
\newcommand {\eea}{\end{equation}}
\newcommand {\C}{\mathbb C} 
\newcommand {\Z}{\mathbb Z} 
\newcommand {\R}{\mathbb R} 
\newcommand {\I}{\mathbb I} 
\newcommand {\F}{\mathfrak F}
\newcommand {\Y}{\mathfrak Y} 
\newcommand{\beq}{\begin{eqnarray}}
\newcommand{\eeq}{\end{eqnarray}}
\newcommand{\bk}{{\bf k}}

\date{\today}
\title{Spin Wave Theory of Spin $1/2$ XY Model with Ring Exchange on a Triangular Lattice}
\author{Solomon A. Owerre}
\affiliation{$^1$Groupe de physique des particules, D\'epartement de physique,
Universit\'e de Montr\'eal,
C.P. 6128, succ. centre-ville, Montr\'eal, 
Qu\'ebec, Canada, H3C 3J7 }

\begin{abstract}
We present the linear spin wave theory calculation of the superfluid phase of a hard-core boson $J$-$K$ model  with nearest neighbour exchange $J$ and four-particle ring-exchange $K$ at half filling on the triangular lattice, as well as the phase diagrams of the system at zero and finite temperatures. A similar analysis has been done on a square lattice \cite{H}. We find similar behaviour to that of a square lattice but with different spin wave values of the thermodynamic quantities. We  also find that the pure $J$ model ($XY$ model) which has a well known uniform superfluid phase with an ordered parameter $M_x=\left\langle S_i^x\right\rangle\neq 0$ at zero temperature is quickly destroyed by the inclusion of a negative-$K$ ring-exchange interactions, favouring a state with a $\left( 4\pi/3 , 0\right)$ ordering wavevector. We further study the behaviour of the finite-temperature Kosterlitz-Thouless phase transition ($T_{KT}$) in the uniform superfluid phase, by forcing the universal quantum jump condition on the finite-temperature spin wave superfluid density. We find that for $K \textless 0$, the phase boundary monotonically decreases to $T=0$ at $K/J = -4/3$, where a phase transition is expected and $T_{KT}$ decreases rapidly while for positive $K$, $T_{KT}$ reaches a maximum at some $K\neq 0$. It has been shown on a square lattice using quantum Monte Carlo(QMC) simulations that for small $K\textgreater 0$ away from the $XY$ point, the zero-temperature spin stiffness value of the XY model is decreased\cite{F}. Our result seems to agree with  this trend found in QMC simulations for 2D systems.
\end{abstract}
\maketitle

\section{Introduction} The effective studies of continuum field theories have resulted in detailed
 predictions for the low-energy physics of quantum spin systems in two dimensions.
 Spin wave theory can provide us with a rather accurate picture of the low-lying states of quantum spin systems. There are several versions of spin wave theory. The standard spin wave theory is based on Holstein-Primakoff representation \cite{J} which was first applied to the study of Heisenberg model by Anderson \cite{D} and further extended to second order by Kubo \cite{E} and Oguchi \cite{I}.
 
 Spin wave theory was thought to be unsatisfactory in the case of $XY$ model until Gomez-Santos and Joannopoulous \cite{A} showed that by a good choice of quantization axis, one can obtain a good theoretical result for $XY$ model. Since then numerous applications of spin wave theory have been carried out on $XY$ model with different lattice configurations and the results obtained so far are in a good agreement with quantum Monte Carlo simulations(QMC) \cite{C,L}.
 
  Another interesting area is the multiple(ring) spin-exchange models. This model was first introduced to describe the magnetic properties of solid $^{3}$He \cite{O}. It incorporates ring-exchange interactions over plaquettes such as spin $1/2$ four-spin $XY$ ring exchange of the form: 
 
 \begin{equation}\label{eqn1.1}
 H_{K} = -K\sum_{\left\langle ijkl \right\rangle} \left(S_{i}^{+}S_{j}^{-}S_{k}^{+}S_{l}^{-} + S_{i}^{-}S_{j}^{+}S_{k}^{-}S_{l}^{+}\right),
 \end{equation}
 where the summation runs over plaquettes with the indices running counter-clockwise, $i$ and $j$ are nearest neighbours lying opposite to $k$ and $l$.
 
The ring exchange interaction is important in Wigner crystal near the melting density \cite{M}. This model, alone or in competition with pure $XY$ model with nearest neighbour exchange has attracted considerable attention over the years. It has been studied extensively on a square lattice using a stochastic series expansion (SSE) quantum Monte Carlo method \cite{F}.  Also,  a comprehensive theoretical study $\left(\text{spin wave theory}\right)$ has been done on a square lattice \cite{H}. It has been suggested in recent works that models of this form may harbour exotic ground state properties, including de-confined quantum critical points or quantum spin liquid phases. 

In this paper, we shall calculate the spin wave values of the Kosterlitz-Thouless temperature, superfluid density and other low temperature thermodynamical quantities of this model on a triangular lattice. The format of the paper is as follows: In Sec.II, we present the model Hamiltonian. In Sec.III, we apply linear wave theory by choosing our quantization axis along the $x$-direction and use it to diagonalize the Hamiltonian and obtain its energy. In Sec.IV, we analyse the dispersion and plot it for some values of $K/J$. In Sec.V, we explore the zero temperature superfluid density using the diagonalized Hamiltonian and plot it for some values of $K/J$. In Sec.VI, we calculate the finite temperature  superfluid density and the value of the Kosterlitz-Thouless temperature for this model. Finally, in Sec.VII, we make some concluding remarks.
\section{Model}
In this section, we shall present the model Hamiltonian and the mean field theory argument of the J-K model on a triangular lattice. Our model Hamiltonian is given by
\begin{equation}
\begin{split}
H = H_{J} + H_{K} = -J\sum_{\left\langle ij \right\rangle} \left(S_{i}^{+}S_{j}^{-} + S_{i}^{-}S_{j}^{+}\right)\\ -K\sum_{\left\langle ijkl  \right\rangle} \left(S_{i}^{+}S_{j}^{-}S_{k}^{+}S_{l}^{-} + S_{i}^{-}S_{j}^{+}S_{k}^{-}S_{l}^{+}\right).
\end{split}
\label{1}
\end{equation}

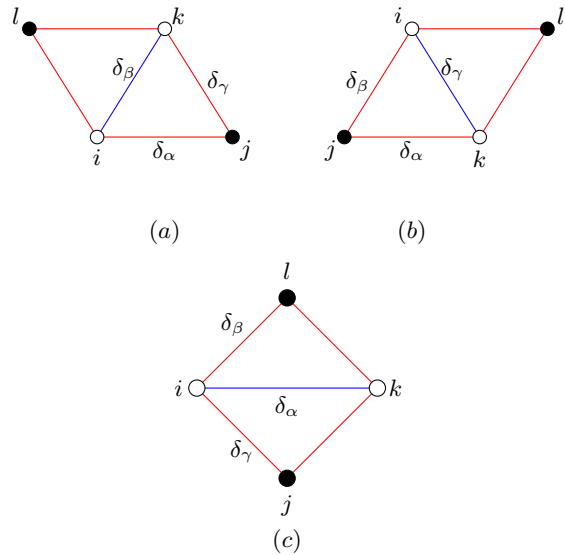
\begin{figure}[h!] 
\centering
\begin{tikzpicture}[scale=1.8] 
\draw[solid,red] (-1,0)--(-2,0) ;
\draw[solid,red] (-2.5,0.8)--(-1.5,0.8);
\draw[solid,red] (-2.5,0.8)--(-2,0);
\draw[solid,red] (-1.5,0.8)--(-1,0);
\draw[solid,blue] (-1.5,0.8)--(-2,0);
\draw[fill= black] (-1,0) circle (0.5mm);
\draw[fill= white] (-2,0) circle (0.5mm);
\draw[fill= white] (-1.5,0.8) circle (0.5mm);
\draw[fill= black] (-2.5,0.8) circle (0.5mm);
\draw(-0.9,-0.1) node[]{$j$};
\draw(-1.4,0.9) node[]{$k$};
\draw(-2,-0.15) node[]{$i$};
\draw(-2.6,0.9) node[]{$l$};
\draw(-1.5,-0.1) node[]{${\delta_{\alpha}}$};
\draw(-1.1,0.4) node[]{${\delta_{\gamma}}$};
\draw(-1.8,0.5) node[]{${\delta_{\beta}}$};
\draw(-1.5,-0.7) node[]{$(a)$};
\end{tikzpicture}
\quad \quad
\begin{tikzpicture}[scale=1.8] 
\draw[solid,red] (1,0)--(2,0) ;
\draw[solid,red] (2.5,0.8)--(1.5,0.8);
\draw[solid,red] (2.5,0.8)--(2,0);
\draw[solid,red] (1.5,0.8)--(1,0);
\draw[solid,blue] (1.5,0.8)--(2,0);
\draw[fill= black] (1,0) circle (0.5mm);
\draw[fill= white] (2,0) circle (0.5mm);
\draw[fill= white] (1.5,0.8) circle (0.5mm);
\draw[fill= black] (2.5,0.8) circle (0.5mm);
\draw(0.9,-0.1) node[]{$j$};
\draw(1.4,0.9) node[]{$i$};
\draw(2,-0.16) node[]{$k$};
\draw(2.6,0.9) node[]{$l$};
\draw(1.5,-0.1) node[]{${\delta_{\alpha}}$};
\draw(1.1,0.4) node[]{${\delta_{\beta}}$};
\draw(1.8,0.5) node[]{${\delta_{\gamma}}$};
\draw(1.5,-0.7) node[]{$(b)$};
\end{tikzpicture}
\quad \quad
\begin{tikzpicture}[scale=1.2]

 \draw[solid,red][->](1,0)--(0,1) ;
 \draw[solid,red] (0,1)--(-1,0);
 \draw[solid,red] (-1,0)--(0,-1);
 \draw[solid,red] (0,-1)--(1,0);
 \draw[solid,blue] (1,0)--(-1,0);
\draw[fill= white] (1,0) circle (0.9mm);
\draw[fill= white] (-1,0) circle (0.9mm);
\draw[fill= black] (0,1) circle (0.9mm);
\draw[fill= black] (0,-1) circle (0.9mm);
\draw(1.2,0 )node[]{$k$};
\draw(-1.2,0) node[]{$i$};
\draw(0,1.3) node[]{$l$};
\draw(0,-1.3) node[]{$j$};
\draw(0,-0.2) node[]{${\delta_{\alpha}}$};
\draw(-0.5,-0.7) node[]{${\delta_{\gamma}}$};
\draw(-0.6, 0.7) node[]{${\delta_{\beta}}$};
\draw(0,-1.7) node[]{$(c)$};
\end{tikzpicture}
\caption{(Color online) The three plaquette orientations $(a)$, $(b)$, and $(c)$ on a triangular lattice with the position coordinates $\delta_{\alpha}= \left(1,0\right)$,              $\delta_{\beta}= \left(-\frac{1}{2},-\frac{\sqrt{3}}{2}\right)$, $\delta_{\gamma}=  \left(-\frac{1}{2},\frac{\sqrt{3}}{2}\right)$.} \label{fig3.2}
\end{figure}

The first summation is over nearest neighbour pairs on a triangular lattice and the second summation runs over the three possible plaquette orientations on a triangular lattice. The subscripts $i$ and $j$ are nearest-neighbours lying opposite to $k$ and $l$ in three different orientations, which together form the three triangular plaquettes Fig.\eqref{fig3.2}. Spin wave theory of this model has been done on a square lattice \cite{H}. Using this model Hamiltonian, we shall reproduce similar plots to that of a square lattice but with different spin wave values of the thermodynamic quantities. It has been shown that this Hamiltonian undergoes a Kosterlitz-Thouless phase transition for $K=0$ at $T_{KT} \approx 0.69$ for 2D model and a superfluid phase for temperatures less than $T_{TK}$ \cite{P}. For $J, K>0$, the Hamiltonian  leads to an in-plane quantum ferromagnet. For $K<0$, there is a sign problem which prevents (QMC) simulations \cite{F} though it is not possible to capture this sign problem in the linear spin wave theory. QMC simulation on a square lattice has been performed for $K>0$ in which there is no sign problem\cite{Q}.
 
 Let us define the spins as classical vectors by making the transformation $S_i^{-}=\rho e^{i\phi_i}$, then the model Hamiltonian becomes
\begin{equation}
\begin{split}
H &= -2J\sum_{\left\langle ij \right\rangle} \rho^2\cos\left(\phi_{i}-\phi_{j}\right)\\&
-2K\sum_{\left\langle ijkl  \right\rangle}\rho^4
\cos\left(\phi_{i}-\phi_{j} +\phi_{k}-\phi_{l}\right).
\end{split}
\label{2}
\end{equation}

Now consider the case $J,K >0$, in this case, minimizing the energy we have $\phi_{i}=\phi_{j}$ for the $J$ term and $\phi_{i}=\phi_{j}$, $\phi_{k}=\phi_{l}$ for the $K$ term. This leads to a ferromagnetic ordering of spins. Consider the case $J,K \textless 0$, in this case, minimizing the energy gives $\phi_{i}-\phi_{j} =\pi$ or $\phi_{i}=0,\phi_{j} =\pi$ for the $J$ term and $\phi_{i}-\phi_{j} +\phi_{k}-\phi_{l}=0$ or $\phi_{i}=\phi_{j} =\phi_{k}=0, \phi_{l}=\pi$ for the $K$ term which leads to ud (up-down) state for the $J$ term and uuud state on the plaquettes. Basically, we have two configurations of spins on the lattice.

 L. Balents and A. Paramekanti \cite{N} considered this model on a triangular lattice with the inclusion of the repulsive interaction $U$  between bosons and also in the regime where $J<<K$. They showed that when $J =0$, the four-spin exchange leads to a manifold of ground states with gapless excitations and critical power-law correlations. When $J\neq 0$, fluctuations select a four-fold ferrimagnetically ordered ground state with a small spin stiffness (͑superfluid) which breaks the global $U(1)$ and translational  symmetry. However, they did not obtain any value of the Kosterlitz-Thouless temperature and the superfluid density which are the main focus in this paper.

 From their Hamiltonian they argued that with $J=0$, the ground state is independent of the sign of $K$. For non-zero $J$, the sign of K is vital. Finally, for $J<0$ with $K<0$, there is a ferromagnetic phase while  $J>0$ leads to a $\sqrt{3} \times \sqrt{3}$ Neel order which are also the ground states for large $\lvert J/K \rvert$. They concluded that there are no phase transitions at any non-zero $J$ other than the well known phases.
 \section{Linear Spin Wave Theory}
The basic assumption of spin wave theory lies on selecting a classical ground state and determining the fluctuation around it. In other words, one considers quantum fluctuations very close
to an ordered ground state configuration of the system under study. By the usual mapping between spins $S=1/2$ and the hard-core bosons, we can view \eqref{1} as a hard-core boson model. For $J>>K$ or $K=0$, the $T=0$ ground state (in-plane ferromagnet) has an ordered parameter $M_x=\left\langle S_i^x\right\rangle \neq 0$ which breaks the $U(1)$ global rotational symmetry or a superfluid phase in the hard-core boson version\cite{R,S,G,U,V}. One can therefore perform a spin wave expansion around this ordered state configuration by introducing the boson operators $a_i$ and $a_i^{\dagger}$ (which represent the low-energy spin wave excitations out of  $\left\langle S_i^x\right\rangle$) and treat other terms in \eqref{1} as perturbations.

We shall follow the procedure of Gomez-Santos and Joannopoulos\cite{A} and choose our quantization axis along the $x$-direction (instead of $z$-direction). This allows us to write the Holstein-Primakoff representation\cite{J} in the linear spin wave theory as
\begin{equation}
\begin{split}
 S_i^x&= \frac{1}{2}-a_i^{\dagger}a_i,\\
 S_i^y &\approx \frac{1}{2i}\left(a_i^{\dagger}-a_i\right).
\end{split}
\label{3}
\end{equation}
The Hamiltonian \eqref{1} can be diagonalized by following the steps outlined by  R. Schaffer $\textit{et al }$ \cite{H} . Firstly, write \eqref{1} in terms of $S_{j}^{x}$ and $S_{j}^{y}$ using $S_{j}^{\pm}= S_{j}^{x} \pm  i S_{j }^{y}$. Secondly,  substitute \eqref{3} into the resulting equation and keep only the quadratic terms. Thirdly, Fourier transform taking into account the summation over all the three plaquettes for the $K$-term. Finally, diagonalize the Hamiltonian using Bogoliubov transformation. After taking all these steps into consideration, it is easy to show that the diagonalized form of the Hamiltonian is:

\begin{eqnarray}
 \begin{split}
 H &=H_{MF} + \sum_{\bold{k}} \left(\omega_{\bold{k}}- A_{\bold{k}} \right) \\ &+\sum_{\bold{k}}\omega_{\bold{k}}\left(\alpha_{\bold{k}}^{\dagger}\alpha_{\bold{k}} +\alpha_{-\bold{k}}^{\dagger}\alpha_{-\bold{k}}\right). 
 \end{split}
 \label{4}
  \end{eqnarray}

The mean-field energy and the coefficients are totally different from those obtained on a square lattice. They are given by
 \begin{eqnarray}
  H_{MF}= -3\left(\frac{1}{2}JN + \frac{1}{8}KN \right),\\\label{eqn1} 
 \omega_{\bold{k}}= \sqrt{A_{\bold{k}}^{2}-B_{\bold{k}}^{2}},\\\label{eqn2}  
  A_{\bold{k}}= J Q_{\bold{k}} + K R_{\bold{k}},\\ \label{eqn3} 
   B_{\bold{k}}= J S_{\bold{k}} + K T_{\bold{k}},\label{eqn4}
 \end{eqnarray}
   where
 \begin{eqnarray}
 Q_{\bold{k}} =3 \left(1-\frac{\gamma_{\bold{k}}}{2}\right),\\
 \label{eqn3.21}
  S_{\bold{k}} = \frac{3}{2} \gamma_{\bold{k}}, \\
  \label{eqn3.22}
R_{\bold{k}}=3\left\lbrace\frac{1}{2}-\frac{1}{2} \gamma _{\bold{k}}+ \frac{1}{8}\left( \gamma _{\bold{k}}+\bar{\gamma _{\bold{k}}} \right)\right\rbrace,\\
\label{eqn3.23}
T_{\bold{k}}= 3\left\lbrace \frac{1}{2} \gamma _{\bold{k}}- \frac{1}{8}\left( \gamma _{\bold{k}}+\bar{\gamma _{\bold{k}}} \right)\right\rbrace, 
\label{eqn3.25}
 \end{eqnarray}
 and the lattice structure constants are given by
  \begin{equation}\label{eqn1.4}
\begin{split}
 \gamma_{\bold{k}}&= \frac{1}{3}\left( \cos k_{x}+ 2\cos \frac{k_{x}}{2}\cos \frac{\sqrt{3}k_{y}}{2}\right),\\
 \bar{\gamma_{\bold{k}}}&=\frac{1}{3}\left( \cos \sqrt{3}k_{y}+ 2\cos \frac{3k_{x}}{2}\cos \frac{\sqrt{3}k_{y}}{2}\right). 
\end{split}
\end{equation}
We can see that the Hamiltonian \eqref{4} reduces to pure XY model in the limit $K=0$ as expected. Eq.\eqref{4} will be used to analyse some properties of the system such as dispersion, ground state, superfluid density etc., all as a function of $K/J$.
 
\section{Dispersion} 
\begin{figure}[h!]
\centering
\includegraphics[scale=0.20]{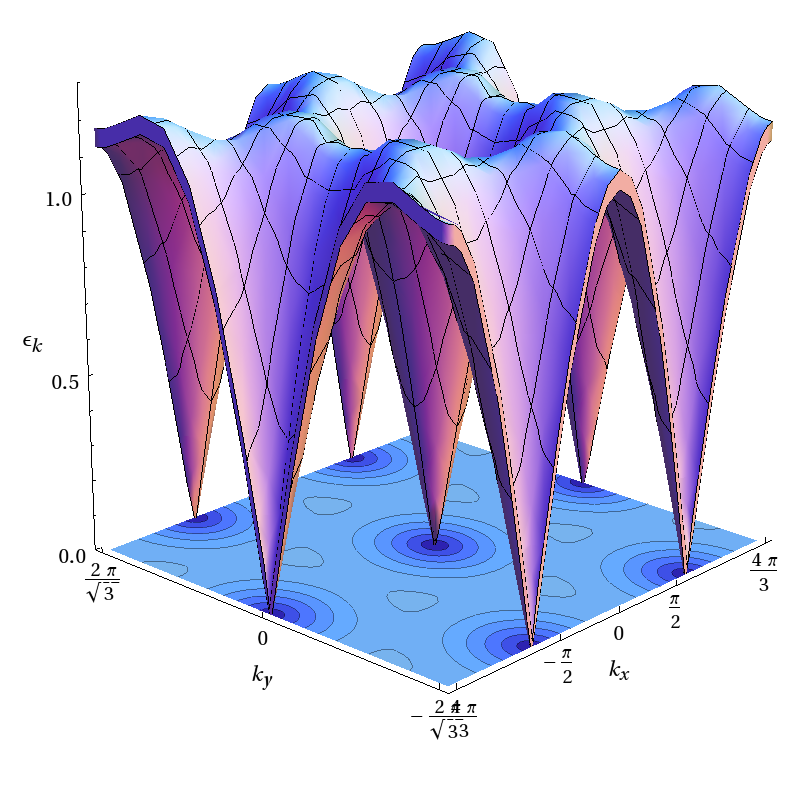} 
\caption{(Color online) The dispersion $\epsilon\left(\bold{k}\right)$ as a function of $\bold{k}$ = $\left(k_x,k_y\right)$ for $K/J$ = $-4/3$  .}\label{fig3.3}
\end{figure}

In this section, we will study the dispersion of $J$-$K$ model as a function of $K/J$. From the diagonalized form of the model Hamiltonian \eqref{4}, we can easily read off the expression for the dispersion relation. For $K/J=-4/3$, it takes the compact form 
 \begin{equation}\label{eqn3.26} 
 \epsilon\left(\bold{k}\right) = 2\sqrt{A_{\bold{k}}^{2}-B_{\bold{k}}^{2}}  =\sqrt{1-\bar{\gamma _{\bold{k}}}},
  \end{equation} 
 where $\bar{\gamma _{\bold{k}}}$ is given by \eqref{eqn1.4}. The plot of \eqref{eqn3.26} is shown in Fig.\eqref{fig3.3}. It shows zero modes at the center and corners of the Brillouin zone. We found that for $K/J=-2$, $\epsilon\left(\bold{k}\right) =\sqrt{0} =0$ for all values of  $\bold{k}$. One cannot obtain any reasonable plot below $K/J=-2$ because the dispersion develops an imaginary part along the $k_x$ and $k_y$ directions. Also the dispersion shows a zero mode at $\bold{k}=(0,0)$ for all values of $K/J$.
 Fig.\eqref{fig3.4} shows the dispersion along the $k_x$ and $k_y$ directions for several values of $K/J$.  Along the $k_x$ direction, the dispersion has  a zero mode at the center of the Brillouin zone $\bold{k}$=$\bold{Q}$=$\left(0,0\right)$  except for  $K/J=-4/3$, where it develops three zero modes (minima) at $\bold{k}$= $\pm \bold{Q}$ =$\left(\pm 4\pi/3 ,0\right)$ and $\bold{k}$= $\bold{Q}$=$\left(0,0\right)$. If one chooses ferromagnetic ordering along the  $k_y$   axis, then the corresponding ordering wave vector is $\bold{k}$= $\pm \bold{Q}$ =$\left(\pm 4\pi/3 ,0\right)$. This is the soft modes of the dispersion for $K/J=-4/3$. The linear spin wave instability of the excitation spectrum at the corner of the Brillioun zone  $\bold{k}$= $\pm \bold{Q}$ =$\left(\pm 4\pi/3 , 0\right)$ occurs for $K/J=-4/3$. For the pure $XY$ model $K/J=0$, there is a gapless excitation at $\bold{Q}$ =$\left(0,0\right)$ (Goldstone mode of the superfluid phase), but there is no minima at the ordering wave vector. Expanding \eqref{eqn3.26} near the zero modes for $K/J= -4/3$, we obtain a linear dispersion relationship:
\bea
 \epsilon(\bold{k}) = \frac{\sqrt{3}}{2}\left(k_{x}^{2}+k_{y}^{2}\right)^{1/2} =  \sqrt{3}J\lvert \bold{k}\rvert. 
 \eea
It follows that the critical velocity is  
 \begin{equation}
 c =  \sqrt{3}J . 
 \end{equation}

\begin{figure}[h!]
\centering
 {\includegraphics[scale=0.25]{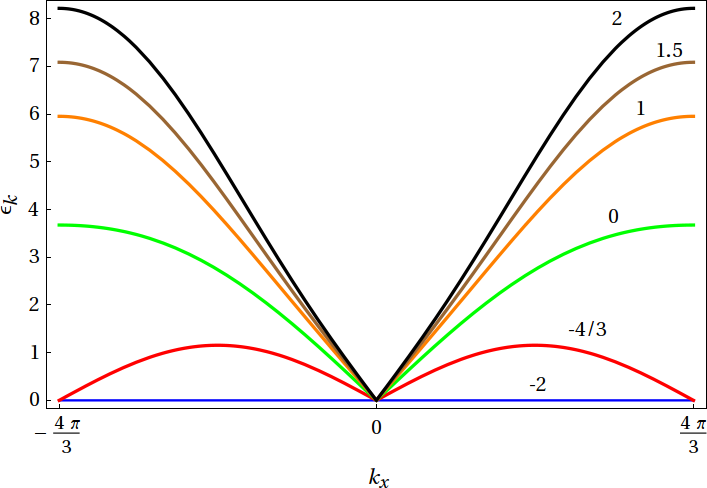}}
 \end{figure}
 \begin{figure}[h!]
 \centering
 {\includegraphics[scale=0.25]{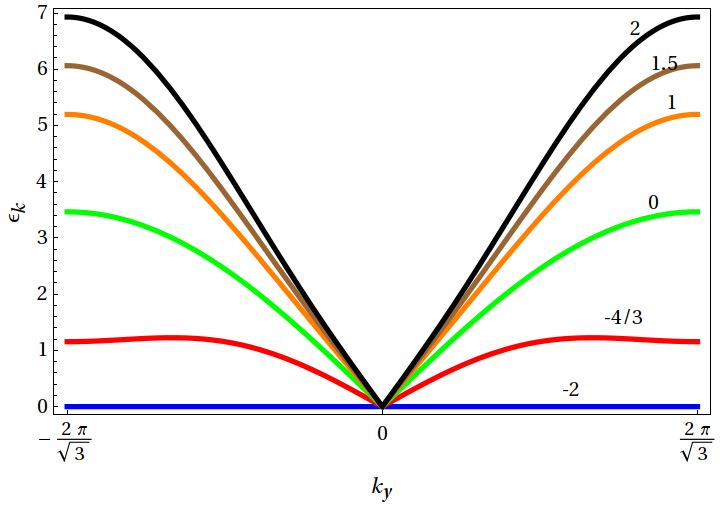} } 
\caption{(Color online) The dispersion along the $\bold{k}$ = $\left(k_x,0\right)$ (top) and $\bold{k}$ = $\left(0,k_y\right)$ (bottom) direction for several values of $K/J$.}\label{fig3.4}
\end{figure}
\section{Superfluid Density at $T=0$}
 In this section, we shall explore the superfluid density of the $J$-$K$ model on a triangular lattice. The ground state superfluid density is usually found by calculating the change in the ground state energy of a spin system with respect to a slow in-plane twist. Consider the rotation of the spin system at site $i$ by an angle $\theta_i$ about the $z$-axis:
\bea
\begin{split}
S_{i}^{+}& \longrightarrow S_{i}^{+}e^{i\theta_i},\\
S_{i}^{-}& \longrightarrow S_{i}^{-}e^{-i\theta_i}.
\end{split}
\eea
Similar transformation applies to all the other sites. Under this rotation, the ring exchange term is found to be 
\begin{equation}\label{eqn3.31}
S_{i}^{+}S_{j}^{-}S_{k}^{+}S_{l}^{-}= S_{i}^{+}S_{j}^{-}S_{k}^{+}S_{l}^{-}e 
^{i(\theta_{i}-\theta_{j}+\theta_{k}-\theta_{l})}.
\end{equation} 
We consider the case of uniform twists, which exist only along the nearest-neighbour bonds oriented along the $x$-axis, so that $\theta_i-\theta_j \equiv \theta_x = \theta$, where $i$ and $j$ are nearest-neighbours along the $x$-direction. Due to the uniformity of the phase twists across the spins in the system, we have $\theta_{i}=\theta_{l}$, $\theta_{j}=\theta_{k}$, using the labelling of Eq.\eqref{1}. Thus, the phase twist angle dependence of the ring exchange terms cancel. The twisted  $J$-$K$ Hamiltonian becomes
 \begin{equation}\label{eqn3.32}
 \begin{split}
 H(\theta) &= -2J\sum_{\left\langle ij \right\rangle}\left\lbrace \left(S_{i}^{ x}S_{j}^{ x}+S_{i}^{y}S_{j}^{y}\right)\cos{\theta} \right.\\&\left. + \left(S_{i}^{ x}S_{j}^{ y} -S_{i}^{y}S_{j}^{x}\right)\sin{\theta}\right\rbrace\\
 &-K\sum_{\left\langle ijkl \right\rangle} \left(S_{i}^{+}S_{j}^{-}S_{k}^{+}S_{l}^{-} + S_{i}^{-}S_{j}^{+}S_{k}^{-}S_{l}^{+}\right).
 \end{split}
 \end{equation}
 In the linear spin wave theory, \eqref{eqn3.32} reduces to
 \begin{equation}\label{3.32}
 \begin{split}
 H(\theta) &= -2J\sum_{\left\langle ij \right\rangle}  \left(S_{i}^{ x}S_{j}^{ x}+S_{i}^{y}S_{j}^{y}\right)\cos{\theta}  \\
 &-K\sum_{\left\langle ijkl \right\rangle} \left(S_{i}^{+}S_{j}^{-}S_{k}^{+}S_{l}^{-} + S_{i}^{-}S_{j}^{+}S_{k}^{-}S_{l}^{+}\right).
 \end{split}
 \end{equation}
Notice that the coefficients of $\sin\theta$ vanishes in the case. Comparing \eqref{1} and \eqref{3.32}, we see that the effect of the twist is to rescale the nearest-neighbour exchange interaction in \eqref{1} as $J \rightarrow J \cos\theta$. Therefore, the diagonalized form of \eqref{3.32} is exactly of the form \eqref{4} with $J$ replaced by  $J \cos\theta$. That is
  \begin{equation}\label{eqn3.33}
  \begin{split}
 H(\theta) =H_{MF}(\theta) + \sum_{\bold{k}} \left(\omega_{\bold{k}}(\theta)- A_{\bold{k}}(\theta) \right) \\ +\sum_{\bold{k}}\omega_{\bold{k}}(\theta)\left(\alpha_{\bold{k}}^{\dagger}\alpha_{\bold{k}} +\alpha_{-\bold{k}}^{\dagger}\alpha_{-\bold{k}}\right),
 \end{split}
 \end{equation}
where the twisted mean-field energy and the coefficients are given by
 \begin{equation}\label{eqn3.34}
  H_{MF}(\theta) = -3\left(\frac{1}{2}JN\cos\theta + \frac{1}{8}KN \right),
 \end{equation}
 \begin{equation}\label{eqn3.35}
  \omega_{\bold{k}}(\theta) = \sqrt{A_{\bold{k}}(\theta)^{2}-B_{\bold{k} }(\theta)^{2}},
 \end{equation}
  \begin{eqnarray}\label{eqn3.36}
  \begin{split}
  A_{\bold{k}}(\theta)&= J Q_{\bold{k}}\cos\theta + K R_{\bold{k}},  \\ B_{\bold{k}}(\theta)&= J S_{\bold{k}}\cos\theta + K T_{\bold{k}},
  \end{split}
 \end{eqnarray}
 the coefficients $ Q_{\bold{k}}, R_{\bold{k}}, S_{\bold{k}},$ and $ T_{\bold{k}}$ remain the same as \eqref{eqn3.21}--\eqref{eqn3.25}.
 
 The free energy is given by
 \begin{equation}\label{eqn3.37}
F (\theta) = -\frac{1}{\beta} \ln Z (\theta) = E_{0}(\theta)+ \frac{1}{\beta}\ln \left(1-e^{-\beta \epsilon _{\bold{k}}(\theta)}\right),
\end{equation}
 where the ground state energy $ E_{0}(\theta)$ is given
  \begin{equation}\label{eqn3.38}
 E_{0}(\theta)= H_{MF}(\theta) + \sum_{\bold{k}} \left(\omega_{\bold{k}}(\theta)- A_{\bold{k}}(\theta) \right).
 \end{equation}
 At zero temperature $\beta = \infty$, the free energy is simply the ground state energy. We shall calculate the superfluid density from the first principle. Taylor expanding the ground state energy we have
  \begin{equation}\label{eqn3.39}
 E_{0}(\theta)= E_{0}(\theta =0) + \frac{1}{2}\rho_{s}\theta ^{2} + O(\theta ^{4}),
 \end{equation}
 
 \begin{equation}\label{eqn3.40}
 \rho _{s}(T=0)=   \frac{1}{N}\frac{\partial ^{2}E_{0}(\theta)}{\partial {\theta ^{2}}}\lvert _{\theta =0}.
 \end{equation}

From \eqref{eqn3.38} and \eqref{eqn3.40} we obtain
 \begin{eqnarray}\label{3.46}
\rho_s(T=0) &=& \frac{3J}{4} + \frac{J}{2N}\sum_{\bk}\Big[Q_{\bk} - \frac{1}{\omega _{\bold{k}}} \\
&\cdot&\big(J\left(Q_{\bold{k}}^{2}- S_{\bold{k}}^{2}\right)+ K\left(Q_{\bold{k}}R_{\bold{k}}- S_{\bold{k}}T_{\bold{k}}\right)\big)\Big] . \nonumber
\end{eqnarray}
 
 Note that we have divided by 2 to account for the dimensionality of the lattice.
 
\begin{figure}[h]
\centering
\includegraphics[scale=0.28]{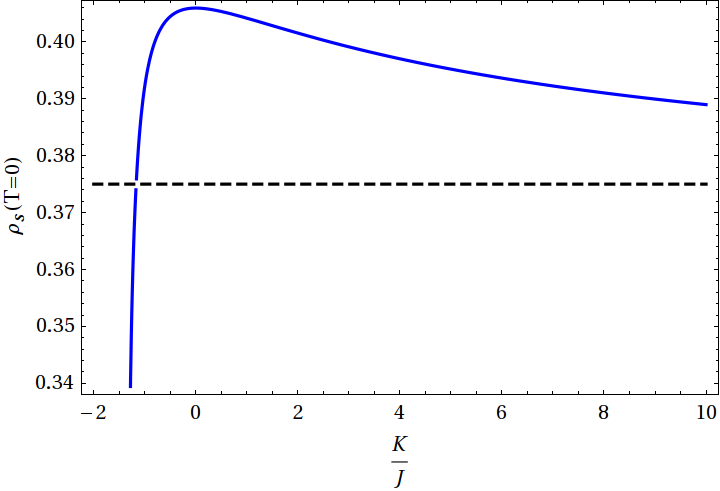}
\caption{(Color online) The superfluid density as a function of $K/J$. The dashed line is the mean-field result, $\rho _{s}^{MF} =0.375$. The linear SW result for $K/J=0$ is $\rho_{s}^{SW}(T=0)=0.4059$.}\label{fig3.7}
\end{figure}

 The plot of $\rho _{s}(T=0)$ is shown in Fig.\eqref{fig3.7} for a range of $K/J$, we have set $J =1/2$. The superfluid density curve has its maximum at $K/J=0$, with a value of $\rho _{s}(T=0)=0.4059$ and  decreases monotonically with increasing $\lvert K \rvert $ as one moves away from this maximum. In other words, the ring exchange term decreases the value of $\rho _{s}(T=0)$ from the pure $XY$ result. On the positive-$K$ side, $\rho _{s}$  decreases relatively gradually, only becomes zero for extremely large values. This is consistent with the result obtained in the dispersion, which indicates that no soft modes develop for moderate values of positive $K$. Similar result was observed on a square lattice \cite{H}, which is consistent with quantum Monte Carlo simulation. On the negative-$K$ side, the value of $\rho _{s}$ decreases rapidly as it approches $K/J=-4/3$.

\section{Superfluid Density and Kosterlitz-Thouless Transition at Finite Temperature}
 In this section, we shall discuss the uniform superfluid phase in the $J-K$ model at finite temperatures. In order to calculate the finite temperature superfluid density, we replace the twisted ground state energy in \eqref{eqn3.40} with the twisted free energy, that is
 \begin{equation}\label{eqn3.47}
  \rho _{s}(T\neq 0)=   \frac{1}{N}\frac{\partial ^{2}F(\theta)}{\partial {\theta ^{2}}}\lvert _{\theta =0}.
 \end{equation}
 Using \eqref{eqn3.37} we obtain
 
  \begin{eqnarray}\label{3.48}
  \rho _{s}(T\neq0)& =&\frac{3J}{4} + \frac{J}{2N}\sum _{\bold{k}}\Big[Q_{\bold{k}}-\frac{1}{\omega _{\bold{k}}}\\
&\cdot&\left(A_{\bold{k}}Q_{\bold{k}}-B_{\bold{k}}S_{\bold{k}}\right)\left(1 + \frac{2}{e^{\epsilon_{\bold{k}}/T}-1}\right)\Big].\nonumber
 \end{eqnarray}
 This expression for $ \rho _{s}(T)$ is plotted as a function of $T/J$ for several values of $K/J$ (see Fig.\eqref{fig3.8}).
 \begin{figure}[h!]
\centering
\includegraphics[scale=0.28]{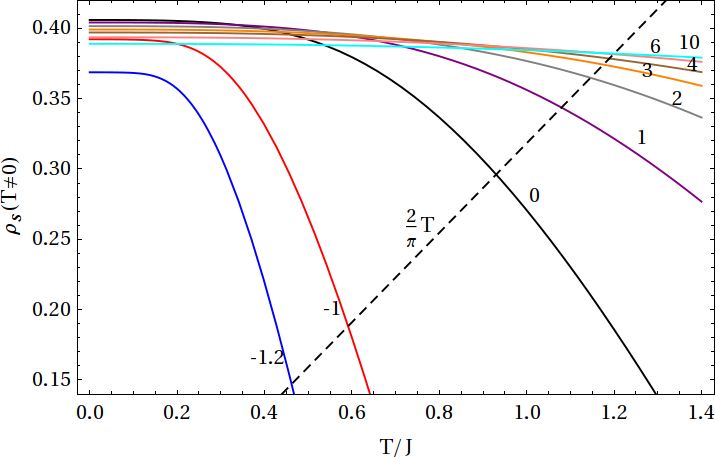}
\end{figure}
\begin{figure}[h!]
\centering
\includegraphics[scale=0.28]{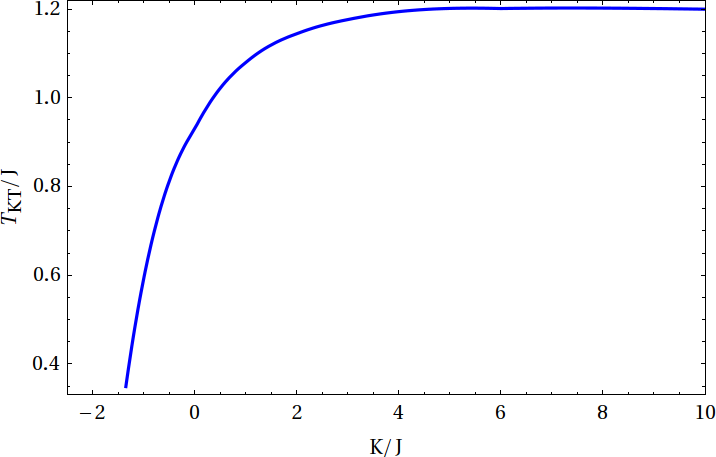}
\caption{ (Color online) The superfluid density as a function of $T/J$. Lines are labelled by the parameter value $K/J$. The dashed line is the universal jump condition (top) and the $KT$ transition phase boundary as calculated from linear $SW$ theory. A maximum of $T_{KT}/J=1.201$ occurs at approximately $K/J=6.019$ (bottom).}\label{fig3.8}
\end{figure}

  The graph shown in Fig.\eqref{fig3.8} is similar to the one obtained on a square lattice\cite{H} which shows slowly decaying superfluid density. The dash line is the so-called $\textit{universal jump}$ condition\cite{T}
\begin{equation}\label{3.49}
\frac{\rho_{s}(T_{KT})}{T_{KT}} = \frac{2}{\pi},
\end{equation}
which accounts for the discontinuity in $\rho_{s}(T)$. The estimate of $T_{KT}$ can be found by solving  $\rho_{s}(T)/{T} =  2/ \pi $, using $\rho_{s}(T)$ from our spin wave theory. Using this procedure we found the $KT$ transition temperature at $T_{KT}=0.9295J$ for $K=0$ and $J=1/2$, the parameter values for the pure $XY$ model. This is shown in Fig.\eqref{fig3.8}, where the dashed line crosses the curve for $K/J=0$. The plot of $T_{KT}$ for non-zero values of $K/J$ is also shown in Fig.\eqref{fig3.8}. It is interesting to see that the maximum does not occur at $K/J = 0$ but at $K/J\approx 6.019$ before  dropping slowly. Similar to the case of square lattice \cite{H}, the $KT$ transition drops to zero for extremely large value of $K/J$  which cannot be captured by a simple spin wave theory. On the negative-$K$ side, the phase boundary drops rapidly as $K/J$ approaches $-4/3$.
\section{Conclusion}
We have presented a comprehensive study of linear spin wave theory of hard-core bosons (zero field $XY$ model) at half filling on the triangular lattice. We studied through linear spin wave theory the destruction of uniform superfluid phase in the bosonic ring exchange model on a triangular lattice at half filling. The dispersion of this model was calculated by applying the traditional Holstein-Primakoff representation and summing over the three plaquettes orientations on a triangular lattice. One might argue that the three plaquettes on the triangular lattice are equivalent. Thus, the effect of the ring exchange term will be the same with that of the square lattice. However, the present calculation showed a spin wave instability and a development of three minima at $\bold{k}$= $\pm \bold{Q}$ =$\left(\pm  4\pi/3 ,0\right)$, and $\bold{k} = \bold{Q}$= $\left(0,0\right)$ for $K/J=-4/3$. One should expect a phase transition from a superfluid phase to another phase at this wave vector $\bold{k}$= $\bold{Q}$ =$\left( 4\pi/3 ,0\right)$ for $K/J=-4/3$. A more careful analysis of quantum Monte Carlo data should provide further insight into this issue. 

Also, the mean field superfluid density and ground state (zero temperature) spin wave superfluid density obtained in this model for $K/J=0$ is bigger than that of a square lattice \cite{H} despite the fact that both are $2D$ systems. This might be due to a larger number of nearest neighbours and plaquettes on a triangular lattice.  We calculated the finite temperature uniform superfluid density and use it to estimate the Kosterlitz-Thouless transition temperature by forcing it to obey the universal quantum jump condition. The maximum value of the Kosterlitz-Thouless transition temperatures was found to occur at $K/J\approx 6.019$. We found that for $K \textless 0$, the phase boundary monotonically decreases to $T=0$ at $K/J = -4/3$, where a phase transition is expected. It has been shown with the  model Hamiltonian \eqref{1} on a square lattice using QMC simulations that for small $K \textgreater 0$ away from the $XY$ point, the zero-temperature spin stiffness value of the XY model is decreased\cite{F}. Our results above seem to agree with  this trend found in QMC simulations.  The effect of the ring exchange term maybe the same on both lattices (square and triangle), but the spin wave values obtained for the thermodynamic quantities are different.

\section*{ACKNOWLEDGEMENTS}
I'm indebted to Roger G. Melko for enlightening discussions. 
Also I would like to thank Akosa Collins and Denis Dalidovich for their encouragement.


\end{document}